\documentclass[sigconf,screen]{acmart}
\AtBeginDocument{%
  }

\copyrightyear{2025}
\acmYear{2025}
\setcopyright{acmlicensed}\acmConference[FSE Companion '25]{33rd ACM
International Conference on the Foundations of Software Engineering}{June
23--28, 2025}{Trondheim, Norway}
\acmBooktitle{33rd ACM International Conference on the Foundations of
Software Engineering (FSE Companion '25), June 23--28, 2025, Trondheim,
Norway}
\acmDOI{10.1145/3696630.3728566}
\acmISBN{979-8-4007-1276-0/2025/06}

\usepackage{overpic}
\usepackage{xspace}
\usepackage{enumitem}
\usepackage{multirow}
\usepackage{mdframed}

\usepackage[utf8]{inputenc} 
\usepackage[T1]{fontenc}    
\usepackage{url}            
\usepackage{booktabs}       
\usepackage{amsfonts}       
\usepackage{nicefrac}       
\usepackage{microtype}      
\usepackage{xcolor}         
\usepackage{nameref}
\usepackage{soul}
\usepackage{mdframed}

\definecolor{royalblue}{RGB}{65,105,225} 
\usepackage{xspace}
\makeatletter
\DeclareRobustCommand\onedot{\futurelet\@let@token\@onedot}
\def\@onedot{\ifx\@let@token.\else.\null\fi\xspace}

\def\etal{\emph{et al}\onedot}

\makeatother

\newcommand{\responseline}{}

\definecolor{lightgray}{HTML}{eeeeee}
\definecolor{darkgray}{HTML}{d8d8d8}
\definecolor{tablecellgreen}{HTML}{d0e6d0}

\definecolor{highlightColor}{rgb}{1, 0.8, 0.6}

\definecolor{amii_magenta}{HTML}{bf477c}
\definecolor{amii_summer}{HTML}{ffcccc}
\definecolor{amii_mustard}{HTML}{faa53c}
\definecolor{amii_sky}{HTML}{6c98ab}
\definecolor{amii_emerald}{HTML}{006c65}
\definecolor{amii_night}{HTML}{003f58}

\definecolor{top1Color}{HTML}{800080}
\definecolor{top2Color}{HTML}{002fa7}

\newcommand{\toponetext}[1]{\textcolor{top1Color}{\textbf{#1}}}
\newcommand{\toptwotext}[1]{\textcolor{top2Color}{\textbf{#1}}}

\newcommand{\codellama}{Code Llama}
\newcommand{\starcoder}{StarCoder2}
\newcommand{\qwen}{Qwen2.5-Coder}

\newcommand{\translation}{Code Lingua}
\newcommand{\edit}{EditEval}
\newcommand{\repair}{Defects4j}

\def\lbl{\textit{LookBack Lens}}
\def\probing{\textit{Probing Classifier}}
\def\uncertainty{\textit{Uncertainty}}
\def\our{\textit{PtTrust}}

\usepackage{subcaption}
\usepackage{soul}

\usepackage[labelfont=bf, textfont=normalfont]{caption}

\newif\ifdisplaycontent
\displaycontenttrue  

\definecolor{color1}{HTML}{E5E5E1}
\definecolor{color2}{HTML}{E0DDEF}
\definecolor{color3}{HTML}{F3D7CA}
\definecolor{color4}{HTML}{E7E2B6}
\definecolor{color5}{HTML}{EAD6FA}

\newcommand{\circled}[1]{{\large \textcircled{\footnotesize #1}}}

\newcommand{\answerYes}[1]{\textcolor{blue}{[Yes]}}
\newcommand{\answerNo}[1]{\textcolor{blue}{[No]}}
\newcommand{\answerNA}[1]{\textcolor{blue}{[NA]}}
\newcommand{\answerPartial}[1]{\textcolor{blue}{[Partial]}}

\newcounter{finding}
\newenvironment{finding}
{
    \refstepcounter{finding}
	\begin{mdframed}[
    	nobreak=true,
    	linecolor=black,
    	roundcorner=12pt,
    	backgroundcolor=gray!05,
    	linewidth=0.5pt,
    	leftmargin=1pt,
    	rightmargin=1pt,
        innerleftmargin=5pt,
        innerrightmargin=5pt,
    	topline=true,
    	bottomline=true,
    	skipabove=10pt
	]
    \textbf{Finding \arabic{finding}}: 
}
{
    \end{mdframed}
    \vspace{3pt}
}

\begin{document}

\title{Risk Assessment Framework for Code LLMs via Leveraging Internal States}



\author{Yuheng Huang}
\affiliation{
  \institution{The University of Tokyo}
  \city{}
  \country{}
}
\email{yuhenghuang42@g.ecc.u-tokyo.ac.jp}

\author{Lei Ma}
\affiliation{
  \institution{The University of Tokyo}
  \city{}
  \country{}
}
\affiliation{
  \institution{University of Alberta}
  \city{}
  \country{}
}
\email{ma.lei@acm.org}

\author{Keizaburo Nishikino}
\affiliation{%
  \institution{Fujitsu Limited}
  \city{}
  \country{}}
\email{nishikino@fujitsu.com}

\author{Takumi Akazaki}
\affiliation{%
  \institution{Fujitsu Limited}
  \city{}
  \country{}}
\email{akazaki.takumi@fujitsu.com}

\renewcommand{\shortauthors}{Huang et al.}

\begin{abstract}
    The pre-training paradigm plays a key role in the success of Large Language Models (LLMs),  which have been recognized as one of the most significant advancements of AI recently. Building on these breakthroughs, code LLMs with advanced coding capabilities bring huge impacts on software engineering, showing the tendency to become an essential part of developers' daily routines. However, the current code LLMs still face serious challenges related to trustworthiness, as they can generate incorrect, insecure, or unreliable code. 
    Recent exploratory studies find that it can be promising to detect such risky outputs by analyzing LLMs’ internal states, akin to how the human brain unconsciously recognizes its own mistakes. Yet, most of these approaches are limited to narrow sub-domains of LLM operations and fall short of achieving industry-level scalability and practicability. To address these challenges, in this paper, we propose {\our}, a two-stage risk assessment framework for code LLM based on internal state pre-training, designed to integrate seamlessly with the existing infrastructure of software companies. The core idea is that the risk assessment framework could also undergo a pre-training process similar to LLMs. Specifically, {\our} first performs unsupervised pre-training on large-scale unlabeled source code to learn general representations of LLM states. Then, it uses a small, labeled dataset to train a risk predictor. We demonstrate the effectiveness of {\our} through fine-grained, code line-level risk assessment and demonstrate that it generalizes across tasks and different programming languages. Further experiments also reveal that {\our} provides highly intuitive and interpretable features, fostering greater user trust. We believe {\our} makes a promising step toward scalable and trustworthy assurance for code LLMs.
\end{abstract}




\maketitle

\section{Introduction}

\begin{mdframed}[backgroundcolor=gray!10,linewidth=0pt,innerleftmargin=10pt,innerrightmargin=10pt]
The biggest lesson that can be read from 70 years of AI research is that general methods that leverage computation are \textbf{ultimately the most effective, and by a large margin.}

\vspace{4pt}

\noindent Seeking an improvement that makes a difference in the shorter term, researchers seek to leverage their human knowledge of the domain, but the only thing that matters in the long run is the \textbf{leveraging of computation.}

\vspace{2pt} 
\hfill \textit{{The Bitter Lesson}}\\
\hfill \textit{— Richard S. Sutton}
\end{mdframed}

\vspace{4pt}


The recent advancements in Large Language Models (LLMs) are continuously exhibiting significant impacts on software engineering, which are increasingly integrated into various stages of the development lifecycle and are becoming an essential component of many software development processes and ecosystems.
In particular, some recent state-of-the-art (SOTA) LLMs demonstrate remarkable performance across a wide range of software engineering tasks, including requirements engineering, software design, development, program analysis, quality assurance, and maintenance~\cite{hou2024large}. For example, a survey from GitHub conducted in late 2023 revealed that 92\% of U.S.-based developers were already utilizing AI coding tools~\cite{github_ai_impact_survey}. This trend continues to grow, highlighting the significant impact of these technologies on the software development industry. As an active industrial contributor in these advancements, Fujitsu has recently developed several pre-trained models, including its most advanced version, Takane~\cite{fujitsu2024takane}, together with highly customized and secure services for private LLM deployments~\cite{fujitsu2024generativeAI}, towards addressing the industrial urgent demand.

Despite the promise, in practice, the current LLMs still face significant challenges in trustworthiness. Recent studies revealed that their generated code often lacks robustness, security, usability, and correctness~\cite{arakelyan2023exploring, pearce2022asleep, barke2023grounded, chen2021evaluating, nguyen2022empirical, jesse2023large, wang2024large}. Additionally, a user study~\cite{tang2024study} found that LLM-generated code introduces atypical errors with patterns distinct from human errors, potentially increasing the cognitive load for developers. These trustworthiness concerns could potentially impede the widespread adoption of LLM technologies. Recognizing this, Fujitsu is actively engaged in R\&D efforts to create more trustworthy LLMs, contributing to advancing reliable, safe, and secure LLM systems.

A practical approach to enhance user trust in LLMs is designing automated tools that help users understand inherent risks and identify potential issues. While compiling test cases to evaluate different aspects (e.g., security, correctness) of the generated code is a viable option, manual test cases can introduce additional labor costs, and LLM-generated tests risk ``chasing one's own tail'' issues. Another approach is to examine the internal workings of LLMs and identify potential risk patterns within their vast network of neurons. This is akin to identifying error neuron behaviors in the human brain, which subconsciously signal when something is amiss~\cite{frank2005error}. Preliminary studies have shown the potential of this approach in natural language processing (NLP) tasks for LLMs~\cite{azaria2023the}.

Although promising, these techniques primarily focus on NLP tasks rather than code-related tasks and have yet to achieve industry-level practicability. A significant limitation is that most existing approaches still follow a pre-LLM paradigm, training classifiers based on LLMs’ internal states using a small domain-specific dataset. This approach often fails to generalize effectively to other tasks, which is a critical drawback given LLMs' foundational and multi-task nature.

To address such challenges, in this paper, we propose {\our}, a two-stage pre-training framework for risk assessment of code LLMs: (1) The first stage leverages unsupervised learning with a recently introduced architecture called Sparse Autoencoders (SAE)~\cite{gao2024scaling}, where OpenAI claims to find scaling laws on them, (2) The second semantic binding stage uses supervised learning to enable generalization across different trustworthiness properties. In particular, {\our} is designed to leverage the existing infrastructure of software companies, allowing it to be seamlessly integrated into existing code LLM service platforms.

We demonstrate the promising of {\our} through a fine-grained, code line-level risk assessment task. This task is based on the observation that when code LLMs make mistakes, it is often that not all code lines are incorrect, but identifying the problematic lines can be time-consuming for developers. By prioritizing incorrect code lines, developers can save significant time and effort. While the company regulations prevent us from directly sharing internal data and results in this paper, we conducted a series of experiments using three representative LLMs with advanced coding capabilities (i.e., {\codellama}, {\starcoder}, {\qwen}) on publicly available datasets. Through only labels on Python languages for code generation and repair datasets (i.e., HumanEval~\cite{chen2021evaluating}, HumanEvalPack~\cite{muennighoff2023octopack}, and QuixBug~\cite{lin2017quixbugs}), {\our} shows highly promising results across tasks (in Python and Java), e.g., including code editing, code translation, and code repair. The results highlight {\our}’s SOTA cross-language and cross-task capabilities in identifying risks associated with code LLMs.


The main contributions of this paper are summarized as follows:
\begin{itemize}[leftmargin=*]
    \item We proposed a two-stage pre-training risk assessment platform {\our} for code LLMs, which is designed to integrate seamlessly with the infrastructure of software companies. This platform is built to scale efficiently to large-scale corpora, potentially providing trustworthy guardrails to the company's code assets generated by LLMs.

    \item We proposed and contributed to establishing a new task, fine-grained code line-level identification, aimed at enhancing developers’ efficiency when interacting with LLMs. This could be potentially useful to enable further research in this direction. 

    \item We systematically evaluated {\our}'s performance across various software engineering tasks and found that it demonstrates strong cross-language and cross-task capabilities attributed to the proposed pre-training technique.

    \item We further demonstrated that {\our} provides interpretable features, enabling users to better understand and interpret LLMs' internal states. This can 
    potentially facilitate many human-centered, interactive, and explainable techniques to enable various trustworthy assurance activities to build trust for code LLMs
    
\end{itemize}

The implementation (e.g., for LLM internal state extraction, analysis, model training, evaluation toolchain) and the labeled datasets contributed in this paper are available at: \responseline{\url{https://github.com/YuhengHuang42/trust_codeLLM}}


\section{Related Work}

\subsection{Internal State Analysis of LLMs}



The rapid advancement of LLM technology has drawn increasing attention from researchers seeking to understand their underlying working mechanisms. Such studies are crucial in improving the performance, trustworthiness, and safety of these models. Notably, one branch of research draws parallels between LLMs and human brains, exploring the relationship between their behavioral characteristics and internal states of LLMs~\cite{mischler2024contextual}. Just as human brains exhibit identifiable patterns when making errors, recent findings indicate that similar patterns can be observed within LLMs. A pioneering work demonstrated that a simple classifier could be trained to uncover some relationships between a model's truthfulness and internal neuron activations~\cite{azaria2023the}. 

Based on these insights, subsequent research makes various exploratory attempts to investigate the connections between correctness~\cite{he2024llm, marks2024the}, safety~\cite{song2024luna, chen2024finding}, privacy concerns~\cite{wu-etal-2024-mitigating-privacy}, toxicity~\cite{liu2024efficient}, and their internal states. Among these studies, an important work by Zhou et al.~\cite{zou2023representation} introduced an approach called representation engineering, which conducts analysis and intervenes in the activations of LLM internal neurons to detect and control their risky behaviors. 

However, one particular drawback of existing approaches is that they rely on the high dimensionality of LLMs' internal states, often involving thousands of neurons per token~\cite{song2024luna}. From a performance perspective, this can lead to the curse of dimensionality~\cite{clarkson1994algorithm}, severely impacting effectiveness. From a computation standpoint, it significantly increases significant complexity in both space and time, posing challenges for real-world deployment. Recently, sparse autoencoders (SAE)~\cite{bricken2023monosemanticity, huben2024sparse} have emerged as a powerful solution to reduce dimensionality and extract interpretable features. OpenAI's recent study ~\cite{gao2024scaling} claims that they found clear scaling laws for training SAE, paving the way for scalable internal feature extraction.

In addition to approaches centered on neuron activations, another line of research focuses on analyzing the attention mechanism—a core component of LLMs' transformer architecture~\cite{vaswani2017attention, su2024codeart}. Chuang \etal introduced \lbl~\cite{chuang-etal-2024-lookback}, a method designed to detect contextual hallucination in NLP tasks by examining the proportion of attention weights assigned to the input context compared to newly generated tokens. Most of the aforementioned studies primarily address trustworthiness issues in general LLMs applied to NLP tasks rather than identifying subtle errors made by code LLMs in multilingual and multi-task settings, which make our work different from these existing work.


\vspace{-4pt}

\subsection{LLMs for Code-Related Tasks}

While LLMs excel at handling a wide range of tasks in a zero-shot manner, code-related tasks, which heavily rely on advanced reasoning, problem-solving, and general coding understanding abilities, remain challenging for general-purpose models. To overcome these challenges, both industry~\cite{roziere2023code, hui2024qwen2, guo2024deepseek, yu2024wavecoder, team2024codegemma} and academia~\cite{li2023starcoder, lozhkov2024starcoder, wei2024magicoder} are actively working to enhance LLMs’ coding capabilities. Related efforts include curating higher-quality code datasets for pre-training, introducing diverse training objectives such as Fill-in-the-Middle (FIM)~\cite{fried2023incoder}, and employing instruction fine-tuning to improve responses to code-related queries. These models serve as the foundation for current state-of-the-art code intelligent code assistants. Their capabilities can be further enhanced through advanced prompting techniques~\cite{chen2023codet, li2023skcoder, gao2023makes, ding2024cycle, jiang2024self, li2024acecoder}, the integration of diverse programming tools~\cite{wang2024executable}, and access to external databases~\cite{chen2024code, su-etal-2024-evor, liu2024codexgraph, phan2024repohyper}, enabling them to tackle increasingly complex tasks. Such an approach, often referred to as code agents~\cite{jimenez2024swebench}, has gained significant interest in both the research community and the software industry in recent years. The rapid advancement of code LLMs has empowered them to tackle highly complex tasks, such as program analysis~\cite{xie2024resym}.

Despite major breakthroughs, the current code LLMs still suffer from challenges, such as limited robustness~\cite{arakelyan2023exploring}, security~\cite{pearce2022asleep}, usability~\cite{barke2023grounded}, and correctness~\cite{chen2021evaluating, nguyen2022empirical, jesse2023large}. Researchers have tried to examine the shortcomings of code LLMs, constructing hierarchical taxonomies that encompass errors, coding style, and maintainability~\cite{liu2023refining, pan2023understanding, liu2023no}. To delve deeper into the mechanisms underlying these errors, a recent study by Kou \etal~\cite{kou2024large} investigated the relationship between LLM errors and their internal attention states. The study revealed that misalignment between model attention and human attention increases the likelihood of errors. These studies indicate that the current code LLMs are still far from perfect, and there is an urgent need to address their trustworthiness issues. 




\section{Method}

Overall, the primary goal of  {\our} is to enable fine-grained identification of errors at the code line level. Specifically, given an incorrect generated response $A$ by an LLM consisting of $n$ lines, the framework should identify a subset ${E \subseteq A}$, representing the erroneous line set.

To achieve this goal, we first provide a high-level introduction to our method {\our}, by explaining the workflow of our technique. Then, we explain the details of each stage in the method.

\subsection{Overview}

\begin{figure*}[t]
    \centering
    \includegraphics[width=0.9\linewidth]{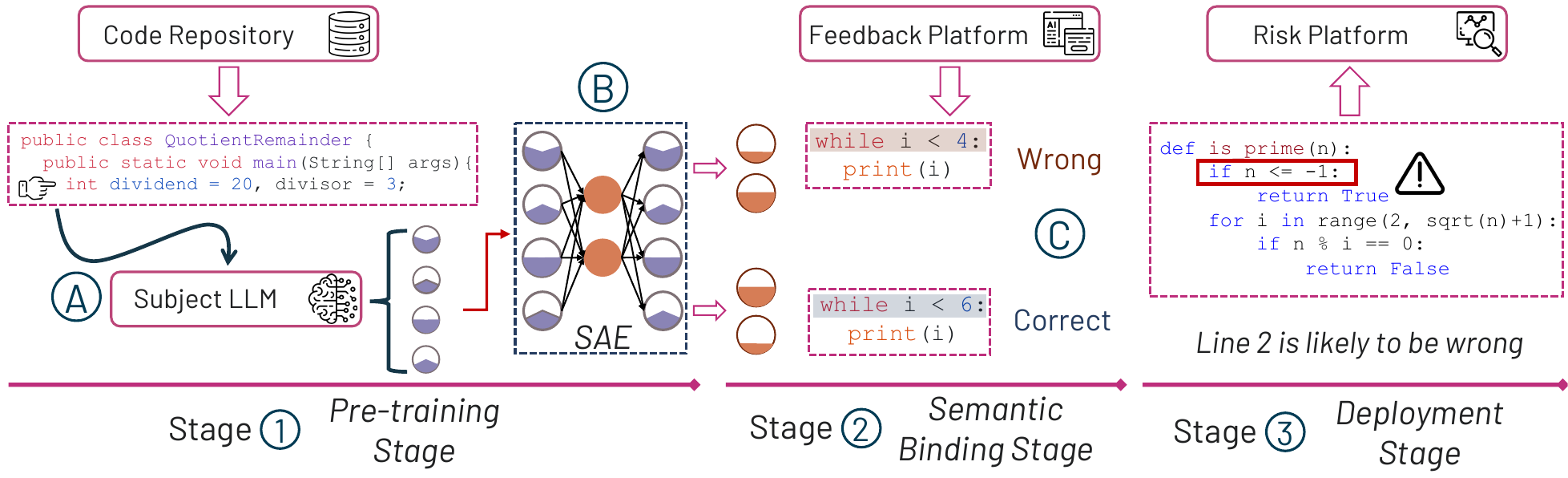}
    \vspace{-8pt}
    \caption{The Summarized Workflow of {\our}. }
    \label{fig:workflow}
    \vspace{-10pt}
\end{figure*}

Overall, three key procedures of {\our} are summarized in Figure~\ref{fig:workflow}. Stage \circled{1} starts with extracting the internal states of a given LLM (\circled{A}) and training a sparse autoencoder (\circled{B}) using these probed states. This stage aims to derive a general and universal low-dimension representation of the LLM’s hidden states, forming the basis for further analysis. This stage operates in a fully unsupervised manner, enabling it to scale effectively to large code corpora. In the software industry, companies often maintain extensive code repositories; this approach offers a cost-effective way to leverage existing software assets. Additionally, depending on the characteristics of the source code, the trained autoencoders can capture features unique to specific companies or even different departments. 

After training the sparse autoencoders, we proceed to Stage \circled{2}. While the abstract representations learned in Stage \circled{1} contain rich information that can help identify risks in code lines, establishing a direct mapping between these representations and specific risks is challenging without external labels. To address this, Stage \circled{2} focuses on collecting fine-grained labels for the correctness of generated code lines. This label collection process can seamlessly integrate into the interaction between LLMs and developers, similar to how ChatGPT incorporates user feedback to improve its models. Developers simply need to highlight the code lines they identify as incorrect during manual review. In addition, as will be demonstrated in our experiments later, even a small number of fine-grained labels (e.g., approximately 100+ code snippets) can significantly outperform existing methods. Combining the representation of autoencoders and the corresponding labels, we can further train a downstream classifier to predict the risks of the generated code (\circled{C}).

Building on the target code LLM (\circled{A}), the sparse autoencoders (\circled{B}), and the downstream classifier (\circled{C}), our framework, {\our}, supports various risk assessment applications in Stage \circled{3}. As an example, we present a prototype for predicting risks at the code line level, though the framework is easily adaptable to other domains by modifying the labels collected in Stage \circled{2}. An additional benefit of {\our} is that the introduced sparse autoencoders can improve the traceability and transparency of the entire system, offering rich internal information besides a single number. Prior research has shown that sparsity is strongly linked to explainability~\cite{elhage2022toy, bricken2023monosemanticity, huben2024sparse}, and we will also demonstrate this advantage in Section~\ref{section:result:rq3}. This enhanced interpretability can significantly increase user trust in a black-box AI system, which could be the key to their experience~\cite{wang2023deepseer}.

Hereafter, we refer to Stage \circled{1} as the pre-training stage and Stage \circled{2} as the semantic-binding stage. In the following sections, we provide more detailed explanations of each stage.

\subsection{Pre-training Stage}
\label{section:method:pretrain}

The pre-training stage is designed to scale efficiently for large-scale code repositories. The first step in this stage is profiling. Specifically, given a text input, we pass it through the target LLM and extract its internal states for each token in the input. We select the early layers (e.g., the first quarter of all network layers) as the profiling target. Previous studies~\cite{huben2024sparse, he2024llm, huang2024active} have shown that early layers are often associated more with high-level decision-making processes, making them a suitable choice for our analysis. Notably, this process involves only a single forward pass without any generation. This is analogous to brain studies, where researchers observe and record brain activity as participants read given materials~\cite{frank2005error, meyer2014psychometric}. Since the ultimate goal is to assess the risks associated with the LLM, it is essential for our ``participant''—the LLM—to encounter incorrect inputs, such as buggy code. Fortunately, extracting commit histories from version control systems provides a straightforward way to collect mistakes and bugs made by developers~\cite{muennighoff2023octopack}. Additionally, we also propose the use of mutation techniques to deliberately generate incorrect code based on existing code snippets. Specifically, we introduce three types of mutators:

\begin{itemize}
    \item \textbf{Switch Inside}: We randomly select and switch two code lines in a given snippet. 
    \item \textbf{Switch Outside}: We randomly select two code snippets from the dataset and swap two lines within each snippet. 
    \item \textbf{Delete Line}: We randomly delete one code line in a code snippet. \footnote{\responseline{The contrastive learning phase requires forming a pair before and after the mutation. For line deletion, the pair consists of the original code line (correct) versus the line following the deleted line (incorrect). Such a setting also applies to the code line error labels in the evaluation stage.}}
\end{itemize}

\responseline{These mutations are simpler to apply in cross-language tasks and have a more significant impact than traditional mutation testing methods, potentially leading to greater differences in neuron behavior.} More fine-grained mutators, such as the universal mutator proposed by Deb et al.~\cite{deb2024syntax}, could also be applied. However, our preliminary experiments did not obtain significant improvements, possibly because the mutations were too fine-grained to have a substantial impact compared to line-level mutations under the scale of our study.
A more systematic investigation into mutation strategies is left for future work.

Given the profiled internal states of LLMs from the text corpus, the next step is to train an autoencoder to extract meaningful information efficiently. Before delving into the technical details, we would like to highlight the motivation behind this step. While it is feasible to directly use the raw internal states for risk assessment, this approach raises two key concerns: practicality and performance. 
From a practical standpoint, the high-dimensional internal neurons of LLMs are both computationally intensive to process and require substantial storage space to handle the data.
For example, {\codellama} includes 8,192 neurons per layer. Generating a single response with 1,000 tokens results in 8,192,000 floating-point values that need to be analyzed. When scaled to multiple responses and users, running such services in real-time becomes prohibitively expensive. From a performance perspective, the large size of the internal states makes it challenging to handle efficiently with machine learning algorithms, especially when the number of labels (e.g., human feedback from Stage \circled{2}) is limited, which could easily lead to over-fitting~\cite{hawkins2004problem}.


To address this challenge, we propose the use of autoencoders~\cite{baldi2012autoencoders} trained on the collected internal states. The autoencoders compress the high-dimensional states into a lower-dimensional latent representation (code), which serves as a more efficient and practical alternative to the original full-dimensional internal states. Their encoder-decoder architecture resembles an hourglass shape with output $\hat{s}$, and they are trained to recover the input states $s$ given the loss function:

\begin{equation}
    \mathcal{L}_{\text{plain}} = || s - \hat{s} ||_{2}^{2}
\end{equation}

In this work, we utilize a variant of autoencoders known as Sparse Autoencoders (SAE)~\cite{ng2011sparse}, which have gained significant traction in Explainable AI (XAI) research on LLMs~\cite{huben2024sparse}. Specifically, we employ TopK autoencoders, a model where OpenAI~\cite{gao2024scaling} has reported observing scaling laws~\footnote{According to the scaling law, the latent dimension of SAE can sometimes exceed the input dimension when trained on large-scale data. However, despite this, the representation remains highly sparse—for instance, only a small subset (e.g., 32 latents) is activated at any given time. This sparsity ensures that the approach remains computationally efficient.}. The encoder is defined as:

\begin{equation}
    z = \text{TopK}(W_{\text{enc}} \cdot (s - b_{\text{pre}}) + b_{\text{enc}})
    \label{eq:plain_loss}
\end{equation}

where $W_{\text{enc}}$ is the encoder weight, $b_{\text{pre}}$ is the pre-encoder bias term~\footnote{This bias is introduced because the SAE operates under dictionary learning context, where we seek to decompose the target activation into a combination of general features. The $b_{\text{pre}}$ is the bias in that decomposition process. Due to space constraints, we cannot provide all the details here and instead refer keen readers to \cite{bricken2023monosemanticity}}, $b_\text{enc}$ is the encoder bias term, and $z$ is the latent (code). $z$ will be the main feature for the following stages of {\our}. TopK is an activation function that retains only the k largest latent values, introducing sparsity by setting all other values to zero. The decoder is simply:

\begin{equation}
    \hat{s} = W_{\text{dec}}\cdot z + b_{\text{pre}}
\end{equation}

where $W_{\text{dec}}$ is the decoder weight. 

The loss described in Eq.~\ref{eq:plain_loss} encourages SAE to use the compressed latent representation $z$ to capture the most salient features necessary for reconstructing the input. However, since our ultimate goal is to assess the risks associated with code LLMs, we aim to make the model more sensitive to error-related information. To achieve this, we incorporate contrastive learning~\cite{chopra2005learning, khosla2020supervised, zhang-etal-2024-truthx, jiang2024novagenerativelanguagemodels}, which encourages the model to distinguish between the representations of correct and incorrect code lines. Specifically, the contrastive loss is defined as:

\begin{equation}
    \mathcal{L}_{\text{cont}}(x_i, x_j, \theta) = \max \left (0, \epsilon - ||f_{\theta}(x_i) - f_{\theta}(x_j) ||_{2}\right)^2  
\end{equation}

where $(x_i, x_j)$ are contrastive pairs, $\theta$ is the parameter of the neural network, and $\epsilon$ defines the lower bound distance between contrastive pairs. In our study, these pairs are created in two ways: by using already-known correct and incorrect code lines or by comparing original code snippets with their mutated counterparts. So the final loss is $\mathcal{L}_{\text{plain}} + \mathcal{L}_{\text{cont}}$.

Finally, an important implementation detail is that, instead of training SAE on every token, we focus exclusively on the newline token (\textbackslash n at the end of every code line). This decision significantly reduces training time and computational costs. For instance, in the pre-training stage, the total number of \textbackslash n tokens is around 280,000, whereas the total number of all tokens is around 9,873,000, about 35 times larger. Moreover, previous studies~\cite{marks2023geometry, jorgensen2023improving, laurito-etal-2024-cluster} have highlighted the difficulty of distilling high-level features (e.g., correctness) from internal states, as these states are entangled with the next-token prediction process. By focusing on only the newline token (\textbackslash n), we can alleviate this issue and mitigate the bias introduced by different tokens.

\subsection{Semantic Binding Stage}

The final stage is semantic binding. While the previous stage produced low-dimension representations $z$ of the code LLMs' internal states for each code line, these representations are not yet linked to the specific risks inherent in the responses. This stage is designed to assign meaning to each vector through a learning process. This is why we refer to this stage as semantic binding: it establishes the connection between the abstract representations and their corresponding risks.

To begin with, we collect a set of coding questions for the target LLM and instruct it to generate responses (which involve hundreds to thousands of inferences). This process differs from Stage \circled{1}, where only a single inference was performed. In Stage \circled{1}, the LLM only \textit{sees}  what is correct or incorrect. However, in Stage \circled{2}, we instruct them to \textit{act} and generate complete responses, and we collect its hidden states throughout the generation process, saving them at the line level. Then, for each generated answer $A_i$ of question $i$, along with its corresponding hidden states $S_i$, we annotate the incorrect code lines $e_j \in E_i$. We transform the hidden states $S_i$ to $Z_i$ using the SAE trained in stage \circled{1}. This results in the formation of a dataset  $(A, Z, E)$ where $A$ is the answer set, $Z$ is the latents obtained from SAE, and $E$ is the identified erroneous lines. 

The next challenge is formulating the mapping $(A, Z) \rightarrow E$ as a machine learning problem. With the SAE parameters frozen, a straightforward approach would be to directly train another DNN for line-level error prediction. However, we found that in practical industrial scenarios, instead of giving verbose predictions of every single code line, it is more natural and intuitive to provide a list of rankings indicating the riskiest code lines to attract developers' attention. By focusing on the most critical lines, this approach reduces cognitive load and makes the risk assessment interface more streamlined and user-friendly. 

Building on this insight, we propose using learning-to-rank (LTR)~\cite{liu2009learning}, a series of well-established techniques in recommender systems that have been widely adopted in industrial applications. Specifically, given a list of code lines $A_i = \{a^j_i | 0 \leq j < n\}$, the goal is to compute a corresponding list of scores $\{r_j | 0 \leq j < n \}$ where each $r_j \in [0, 1]$ represents the risk level associated with the $j$-th code line. A higher score means a more severe risk.

In particular, we employ \textit{NeuralNDCG} loss~\cite{pobrotyn2021neuralndcg} to train the network. This loss function integrates seamlessly into standard DNN training pipelines and serves as a differentiable approximation of sorting, enabling the network to learn to rank code lines effectively based on their risk levels. Due to space limitations, we omit the technical details of the loss function here. During training, we assign a score of 0 to correct code lines. For incorrect code lines, we rank them based on their length under the assumption that longer lines warrant greater attention.

\section{Result and Analysis}
\label{section:result}


\responseline{The first research question (RQ1) addresses the primary focus of this paper: the ability to identify errors at a fine-grained code line level. This ability enables risk assessment tools to highlight the most vulnerable lines of code, allowing users to focus their attention on quickly identifying and addressing potential issues. In contrast, the second research question (RQ2) examines a more traditional aspect: determining whether an entire code snippet is incorrect. While RQ1 is about the relative risk and prioritization of individual code lines, RQ2 addresses the overall correctness of a code segment. Finally, we assess the robustness and interpretability of the features learned by our method, which is essential for fostering user trust.}

\subsection{Experiment Settings}

\subsubsection{Dataset}

Dataset selection is critical for evaluation. Due to company regulations, we cannot release internal datasets. Instead, we use open-source datasets that closely resemble real-world industrial scenarios to ensure the evaluation’s relevance and practicality. 

\textbf{For the pre-training stage}, we select three datasets comprising multilingual programming problems and their solutions. Notice that the data used in this stage can be replaced with code from internal software repositories within a software company. The first dataset is the LeetCode Dataset~\cite{wang2024testeval, leetcode_dataset}, from which we extract approximately 2,358 coding problems across four programming languages. This dataset does not contain buggy code line information, so we apply mutation as described in Section~\ref{section:method:pretrain}. The other two are HumanEvalPack (EvalPack for short)~\cite{muennighoff2023octopack} and DebugBench~\cite{tian2024debugbench}. EvalPack dataset is derived from the original HumanEval~\cite{chen2021evaluating} dataset, with manually introduced bugs added to each of the 164 HumanEval solutions. It spans six programming languages. DebugBench contains 4,253 instances in three programming languages. Each instance is accompanied by detailed annotations of bugs and their corresponding correct implementations. \responseline{Therefore, we can leverage this labeling directly without performing mutations.} These two datasets can be substituted with commit histories from software companies’ version control systems.

\textbf{For the semantic binding stage}, we aim to replicate real-world scenarios where labels are scarce and highly valuable. Furthermore, to further evaluate {\our}'s performance across languages, we intentionally limit this stage to include only Python code. As such, we choose HumanEval~\cite{chen2021evaluating} for Python code generation, EvalPack~\cite{muennighoff2023octopack} and Quixbug~\cite{lin2017quixbugs} for Python code repair. In total, there are 368 instances, which is two orders of magnitude smaller than the dataset used in the pre-training stage.

\textbf{For the evaluation stage}, we select three diverse tasks: code editing, code translation, and code repair. Code editing involves providing LLMs with original code along with new requirements and instructing them to make the corresponding modifications. Code translation requires converting a program from one programming language to another. Code repair focuses on fixing errors in a given code snippet. We select \edit~\cite{li2024instructcoder}, \translation~\cite{pan2024lost} and \repair~\cite{just2014defects4j} dataset for each of the task. Specifically, for \translation, we select the Python-Java AVATAR subset, which contains abundant test cases (i.e., on average, 50 for each instance). For \repair, we follow the settings (e.g., prompts) in~\cite{xia2023automated} and merge \repair-1.2 and \repair-2.0 into one to test LLMs' single function repair ability. 

The next challenge is to obtain fine-grained labels after code LLMs generate answers to the selected questions. For repair tasks (\responseline{e.g., Defects4j}), the modifications are relatively limited compared to the ground truth code, allowing us to use standard \textit{diff} tools to automatically label the LLM-generated incorrect code. However, for HumanEval, {\edit} and {\translation}, there might be several correct alternatives that are syntactically different. Manually labeling all instances is highly time-consuming. To address this, we instruct GPT-4o to identify erroneous code lines. To validate GPT-4o's accuracy in identifying these lines, we also ask it to provide repair solutions based on the identified errors. For those problems that GPT-4o fails to fix, the authors perform a manual review by sampling these cases. The sampling is conducted with a 90\% confidence level and a margin of error of 20\%. The overall statistics are shown in Table~\ref{table:manual_label}. In general, we believe the label quality is reliable. Finally, we make manual adjustments in cases where the provided code lines do not align with the generated answers.

\begin{table*}[htbp]
\resizebox{0.98\linewidth}{!}{
    \begin{tabular}{lccccccccc}
        \toprule
        \multirow{2}{*}{\textbf{Model}} & \multicolumn{3}{c}{\textbf{HumanEval}}                & \multicolumn{3}{c}{\textbf{\translation}}              & \multicolumn{3}{c}{\textbf{\edit}}                 \\
                \cmidrule(lr){2-4} \cmidrule(lr){5-7} \cmidrule(lr){8-10} 
                               & \textbf{LLM Error} & \textbf{GPT4o Error} & \textbf{GPT4o failed} & \textbf{LLM Error} & \textbf{GPT4o Error} & \textbf{GPT4o failed} & \textbf{LLM Error} & \textbf{GPT4o Error} & \textbf{GPT4o failed} \\
        \midrule
        \codellama              & 44/164    & 17/44       & 2/9                & 136/250   & 60/136      & 2/14               & 80/194    & 11/80       & 1/7                \\
        \starcoder              & 66/164    & 25/66       & 2/13               & 130/250   & 69/130      & 4/14               & 79/194    & 10/79       & 1/7                \\
        \qwen                   & 58/164    & 21/58       & 2/10               & 107/250   & 56/107      & 4/14               & 60/194    & 14/60       & 1/8 \\  
        \bottomrule
    \end{tabular}
}
\caption{Labeling statistics for three datasets. \textbf{LLM Error} means the total number of errors generated by the LLM during its responses. \textbf{GPT-4o Error} means the number of cases where GPT-4o's attempted fixes failed to produce the correct code. \textbf{GPT-4o Failed} refers to the number of instances manually confirmed where GPT-4o failed to label the code line errors correctly.}
\vspace{-5mm}
\label{table:manual_label}
\end{table*}

\subsubsection{Model}

When selecting code LLMs for evaluation, we prioritize open-source models with relatively strong capabilities to ensure they can generate reasonable answers for the given datasets. This avoids trivial results where all code lines would be labeled as errors due to poor model performance. Following this guidance, we select the following models: {\codellama}-34B~\cite{roziere2023code} and {\qwen}-32B~\cite{hui2024qwen2}, which are representative of industrial-grade models, as well as {\starcoder}-15B~\cite{lozhkov2024starcoder}, which comes from the academic and open source community.

\subsubsection{Baseline} There has been little research focusing on fine-grained errors made by LLMs. Among the existing approaches: 
\begin{itemize}[leftmargin=*]
    \item {\uncertainty}-based methods are the most straightforward~\cite{huang2023look, spiess2024calibration}, as they directly analyze the model's output. One possible way is to aggregate confidence scores for all tokens within a code line. In this work, we use the \emph{mean} as an indicator of its risk ($1-confidence$). We aggregate at the line level in RQ1 and the code snippet level in RQ2. Despite their simplicity, such methods can perform well in some cases because they do not rely on any training data, thus avoiding associated biases. 

    \item {\lbl} is a recent approach that detects in-context errors by analyzing the ratio of attention weights assigned to the context versus newly generated tokens. By adjusting the window size, {\lbl} enables fine-grained error detection. In their original work, Logistic Regression was used for error detection. To provide a fair comparison and align model complexity with our approach, we also evaluate its performance using a Multilayer Perceptron (MLP)-based implementation.

    \item {\probing}~\cite{azaria2023the, liu2024efficient} refers to the approaches that directly train a classifier based on models' internal states. We use a lightweight 4-layer Multilayer Perceptron (MLP) with a hidden size of 32. This architecture is shared for both {\probing} and {\our} with only the difference in the input layer.
\end{itemize}

\subsubsection{Hardware}

To conduct our large-scale experiments, we utilized a server with AMD 3955WX CPU, 256GB RAM, and four NVIDIA A4000 GPUs (16GB VRAM of each) and two servers each with AMD Pro 5955WX CPU, 256GB RAM, and two NVIDIA A6000 GPUs (48GB VRAM of each).

\subsection{RQ1: How effective is our method in identifying incorrect code lines?}

\begin{table*}[htbp]
    \begin{tabular}{lcccccccccc}
        \toprule
        \multirow{2}{*}{\textbf{Method}} & \multicolumn{3}{c}{\bf \edit} & \multicolumn{3}{c}{\bf \translation} & \multicolumn{3}{c}{\bf \repair} \\
        \cmidrule(lr){2-4} \cmidrule(lr){5-7} \cmidrule(lr){8-10}
        & \textbf{Top-1} & \textbf{Top-3} & \textbf{Top-5} & \textbf{Top-1} & \textbf{Top-3} & \textbf{Top-5} & \textbf{Top-1} & \textbf{Top-3} & \textbf{Top-5} \\
        \midrule

        \lbl-Logistic & 0.100 & 0.347 & 0.587 & 0.051 & 0.142 & 0.252 & 0.109 & 0.262 & 0.395 \\

        \lbl-MLP & 0.091 & 0.346 & 0.561 & 0.051 & 0.145 & 0.232 & 0.110 & 0.277 & 0.405\\ 

        \probing & 0.252 & \toptwotext{0.560} & 0.709 & \toptwotext{0.142} & \toptwotext{0.325} & \toptwotext{0.440} & 0.229 & 0.397 & 0.507  \\

        \uncertainty & \toptwotext{0.256} & 0.557 & \toptwotext{0.722} & 0.040 & 0.152 & 0.268 & \toptwotext{0.262} & \toponetext{0.499} & \toponetext{0.603} \\

        \midrule

        Ours (\our) & \toponetext{0.349} & \toponetext{0.641} & \toponetext{0.798} & \toponetext{0.223} & \toponetext{0.419} & \toponetext{0.528} & \toponetext{0.287} & \toptwotext{0.479} & \toptwotext{0.595} \\

        \bottomrule
    \end{tabular}
    \caption{\emph{Top-K Hit Rate} of {\codellama} in code line error identification. Different colors are used to highlight the \toponetext{best} and \toptwotext{second-best}}
    \label{table:rq1:codellama}
    \vspace{-8mm}
\end{table*}

\begin{table*}[htbp]
    \begin{tabular}{lcccccccccc}
        \toprule
        \multirow{2}{*}{\textbf{Method}} & \multicolumn{3}{c}{\bf \edit} & \multicolumn{3}{c}{\bf \translation} & \multicolumn{3}{c}{\bf \repair} \\
        \cmidrule(lr){2-4} \cmidrule(lr){5-7} \cmidrule(lr){8-10}
        & \textbf{Top-1} & \textbf{Top-3} & \textbf{Top-5} & \textbf{Top-1} & \textbf{Top-3} & \textbf{Top-5} & \textbf{Top-1} & \textbf{Top-3} & \textbf{Top-5} \\
        \midrule

        \lbl-Logistic & 0.136 & 0.405 & 0.585 & 0.074 & 0.192 & 0.256 &  0.122 & 0.301 & 0.426 \\

        \lbl-MLP & 0.189 & 0.484 & 0.628 & \toptwotext{0.113} & \toptwotext{0.196} & \toptwotext{0.293} & 0.149 & 0.317 & 0.430\\ 

        \probing & 0.202 & 0.502 & \toptwotext{0.701} & 0.053 & 0.121 & 0.202 & 0.103 & 0.317 & 0.438 \\

        \uncertainty & \toptwotext{0.239} & \toptwotext{0.514} &0.682 & 0.034 & 0.149 & 0.258 & \toptwotext{0.192} & \toptwotext{0.361} & \toptwotext{0.486} \\

        \midrule

        Ours (\our) & \toponetext{0.302} & \toponetext{0.555} & \toponetext{0.754} & \toponetext{0.216} & \toponetext{0.404} & \toponetext{0.565} & \toponetext{0.207} & \toponetext{0.388} & \toponetext{0.503} \\

        \bottomrule
    \end{tabular}
    \caption{\emph{Top-K Hit Rate} of {\starcoder} in code line error identification. Different colors are used to highlight the \toponetext{best} and \toptwotext{second-best}}
    \label{table:rq1:starcoder}
    \vspace{-8mm}
\end{table*}

\begin{table*}[htbp]
    \begin{tabular}{lcccccccccc}
        \toprule
        \multirow{2}{*}{\textbf{Method}} & \multicolumn{3}{c}{\bf \edit} & \multicolumn{3}{c}{\bf \translation} & \multicolumn{3}{c}{\bf \repair} \\
        \cmidrule(lr){2-4} \cmidrule(lr){5-7} \cmidrule(lr){8-10}
        & \textbf{Top-1} & \textbf{Top-3} & \textbf{Top-5} & \textbf{Top-1} & \textbf{Top-3} & \textbf{Top-5} & \textbf{Top-1} & \textbf{Top-3} & \textbf{Top-5} \\
        \midrule

        \lbl-Logistic & 0.122 & 0.346 & 0.594 & 0.014 & 0.092 & 0.162 & 0.129 & \toptwotext{0.350} & \toptwotext{0.453} \\

        \lbl-MLP & 0.115 & 0.344 & 0.594 & 0.019 & 0.082 & 0.160 & 0.131 & 0.339 & 0.443\\ 

        \probing & 0.214 & 0.399 & 0.559 & \toptwotext{0.051} & 0.116 & 0.229 & 0.124 & 0.289 & 0.401 \\

        \uncertainty & \toptwotext{0.257} & \toptwotext{0.558} & \toptwotext{0.717} & 0.031 & \toptwotext{0.154} & \toptwotext{0.245} & \toponetext{0.254} & \toponetext{0.474} & \toponetext{0.588} \\

        \midrule

        Ours (\our) & \toponetext{0.287} & \toponetext{0.581} & \toponetext{0.728} & \toponetext{0.177} & \toponetext{0.328} & \toponetext{0.434} & \toptwotext{0.193} & 0.328 & 0.429 \\

        \bottomrule
    \end{tabular}
    \caption{\emph{Top-K Hit Rate} of {\qwen} in code line error identification. Different colors are used to highlight the \toponetext{best} and \toptwotext{second-best}}
    \vspace{-8mm}
    \label{table:rq1:qwen}
\end{table*}

The first RQ highlights the primary focus of this work: fine-grained code line error detection. To evaluate each method, we use the metric \emph{Top-K Hit Rate}. Specifically, each method assigns a score to each code line in a given problem. From these scores, we select the Top-K lines with the highest scores and calculate the proportion of buggy tokens within these lines relative to the total number of buggy tokens. We evaluate {\codellama}, {\starcoder}, and {\qwen} on {\edit}, {\translation}, and {\repair} datasets. The results of Top-1, Top-3 and Top-5 Hit Rate are shown in Table~\ref{table:rq1:codellama}-\ref{table:rq1:qwen}.

We can observe that {\our} achieves state-of-the-art performance across all three code LLMs. Specifically, {\our} attains the highest Top-K scores for all tasks on {\starcoder} and for all tasks except {\repair} on {\codellama} and {\qwen}. For {\repair}, the {\uncertainty} method performs surprisingly well. This performance gap might be attributed to the very limited amount of training data related to repair tasks during the semantic binding stage. For instance, {\qwen} made only 58 errors in the EvalPack dataset and two errors in the QuixBug dataset. This restricts the number of error-related training instances to only 60 in total. Additionally, these training data are exclusively in Python, whereas test set Defects4j is in Java, posing a significant challenge for learning-based methods to generalize effectively. A slightly larger and more comprehensive data could potentially improve the performance of {\our} and other training-based methods. 
However, {\our} remains competitive on this task. In contrast, for {\translation} tasks, {\our} significantly outperforms other methods. For instance, on {\codellama}, {\our} identifies 5.5 times as many buggy code lines as the {\uncertainty} method on the Top-1 metric and 5.7 times on {\qwen}. 

On the other hand, {\lbl} appears to struggle to produce good results. First of all, {\lbl} is originally designed for NLP tasks, which may limit its effectiveness in handling code-related tasks. Adapting it to better suit the unique characteristics of code tasks might be necessary. Another possible reason is the significant differences in LLMs' attention patterns between the training data used in the semantic binding stage and the data used during evaluation, making it difficult for downstream models to generalize effectively. Especially considering we are evaluating the methods under cross-task and cross-lingual settings. While the {\probing} performs better than {\lbl}, it still falls short of the out-of-the-box {\uncertainty} method. These findings support our hypothesis that in the era of LLMs, training classifiers for error prediction requires caution. The operational range of LLMs can far exceed the scope of the limited training data, leading to challenges in achieving reliable generalization.

\begin{finding}
\label{finding:1}
    Stakeholders should exercise caution when applying learning-based risk classifiers, especially in scenarios where there is a potential data distribution shift between training and testing, which could often be the case for general-purpose code LLMs. However, thanks to the pre-training stage, {\our} achieves SOTA performance while maintaining robustness across tasks and programming languages.
\end{finding}


\subsection{RQ2: How effective is our method in identifying incorrect code snippets?}

While predicting the correctness of an entire code snippet is not the primary focus of this study, it provides a useful perspective for evaluating the quality of the learned features. Following the settings of previous studies~\cite{azaria2023the, zou2023representation}, we use the hidden state of the final token as the feature to train a simple classifier. We exclude {\lbl} from this RQ since it is not designed for such tasks. Our preliminary results for {\lbl} on {\edit} and {\translation} also reveal that it trivially predicts zero for all test instances, regardless of whether the feature is derived from the final token or the entire generated code snippet. For the {\uncertainty} method, we continue to use mean aggregation but at the code snippet level. A key design choice is the selection of the threshold for determining whether a line is classified as erroneous. To optimize this threshold, we use Youden’s J statistic~\cite{fluss2005estimation}, which balances sensitivity and specificity. The cutoff is determined using the training dataset from the semantic binding stage (i.e., HumanEval, EvalPack, and Quixbug).

The results measured in \responseline{code-snippet level error detection} accuracy are shown in Table~\ref{table:rq3:class}. In general, {\our} continues to deliver the best performance, achieving state-of-the-art results in 7 out of 9 cases. A slightly different observation this time is that the {\uncertainty} method performs the worst. This aligns with findings from previous studies~\cite{huang2023look, spiess2024calibration} indicating that LLMs are generally poorly calibrated in their out-of-the-box state. While uncertainty scores can effectively highlight relative risks between different code lines, they may struggle with snippet-level correctness due to the excessive noise introduced in the process. For example, not all tokens within a snippet contribute equally to the overall correctness, which makes aggregation methods less reliable in such scenarios.

\begin{table*}[htbp]
    \renewcommand{\arraystretch}{1.0}
    \resizebox{0.98\linewidth}{!}{
    \begin{tabular}{lcccccccccc}
        \toprule
        \multirow{2}{*}{\textbf{Method}} & \multicolumn{3}{c}{\bf \edit} & \multicolumn{3}{c}{\bf \translation} & \multicolumn{3}{c}{\bf \repair} \\
        \cmidrule(lr){2-4} \cmidrule(lr){5-7} \cmidrule(lr){8-10}
        & {\small \textbf{\codellama}} & {\small \textbf{\starcoder}} & {\small \textbf{\qwen}} & {\small \textbf{\codellama}} & {\small \textbf{\starcoder}} & {\small \textbf{\qwen}} & {\small \textbf{\codellama}} & {\small \textbf{\starcoder}} & {\small \textbf{\qwen}} \\
        \midrule

        \probing & 0.516 & \toptwotext{0.606} & \toptwotext{0.601} & \toptwotext{0.569} & \toptwotext{0.571} & \toptwotext{0.577} & 0.437 & \toptwotext{0.667} & 0.154 \\

        \uncertainty & \toptwotext{0.552} & 0.565 & 0.586 & \toponetext{0.621} & 0.563 & \toponetext{0.617} & \toptwotext{0.499} & 0.346 & 0.309 \\

        \midrule

        Ours (\our) & \toponetext{0.593} & \toponetext{0.622} & \toponetext{0.627} & 0.508 & \toponetext{0.595} & 0.548 & \toponetext{0.515} & \toponetext{0.845} & \toponetext{0.355} \\

        \bottomrule
    \end{tabular}
    }
    \caption{Classification Performance for three datasets with three code LLMs.}
    \label{table:rq3:class}
    \vspace{-6mm}
\end{table*}

\begin{finding}
\label{finding:2}
    {\our} achieves SOTA performance in entire code snippet error prediction across all baselines. Unlike the results in RQ1, where the {\uncertainty} method showed competitive performance, it struggles with this task.
\end{finding}

\subsection{RQ3: Can our method learn robust and interpretable features?}
\label{section:result:rq3}

In addition to evaluating downstream model performance quantitatively, this section explores {\our}’s effectiveness by analyzing the learned features (latent representations of SAE) from two perspectives: whether the learned features can convey robust representations across programming languages and whether these features are interpretable.

\subsubsection{Cross-language ability}

A key concern with internal-state-based methods is that, theoretically, the internal neurons of LLMs primarily encode information related to next-token prediction, which is their training objective. This makes it highly challenging to extract insights into the model's beliefs, decisions, or other high-level reasoning. Next-token prediction typically emphasizes syntactic features at the time of inference. For instance, experiments by Bricken \etal~\cite{bricken2023monosemanticity} demonstrate that the activated latent representations can effectively distinguish Arabic script from base64 strings.

\begin{figure*}[ht]
    \centering
    \begin{minipage}[b]{0.3\textwidth}
        \centering
        \includegraphics[width=0.65\textwidth]{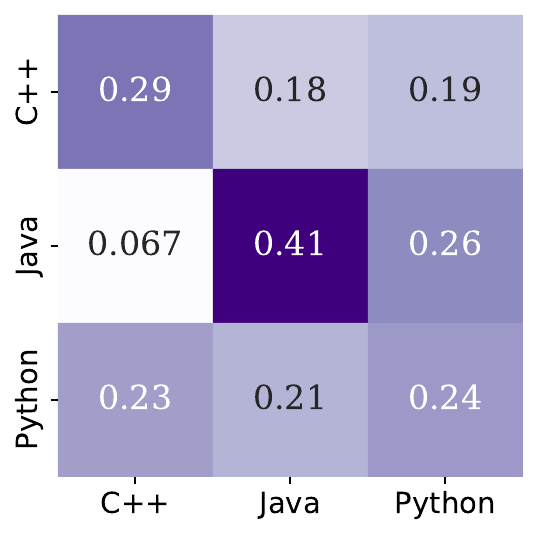}
        \caption*{\codellama}
    \end{minipage}
    \hfill 
    \begin{minipage}[b]{0.3\textwidth}
        \centering
        \includegraphics[width=0.65\textwidth]{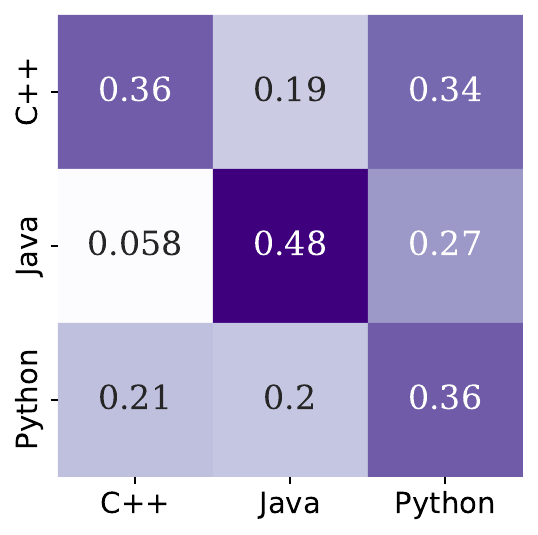} 
        \caption*{\starcoder}
    \end{minipage}
    \hfill 
    \begin{minipage}[b]{0.3\textwidth}
        \centering
        \includegraphics[width=0.65\textwidth]{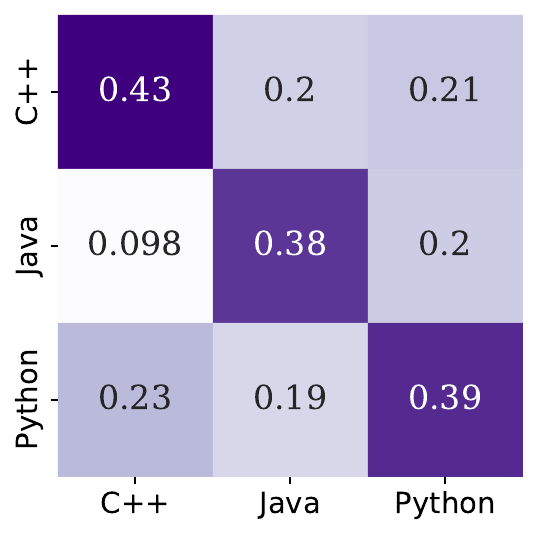} 
        \caption*{\qwen}
    \end{minipage}
    \vspace{-2mm}
    \caption{Distribution difference across tasks for three LLMs. The lower triangle represents the distance between each language pair in the LeetCode dataset, while the upper triangle shows the corresponding distances within the EvalPack dataset. The diagonal entries indicate the cross-task distance (LeetCode v.s. EvalPack) for the same language.}
    \label{fig:rq3:distance}
    \vspace{-2mm}
\end{figure*}

To evaluate whether the SAE in {\our} can learn robust representations for risk assessment, we analyze the activation distribution differences across tasks and programming languages. For this purpose, we use the LeetCode and EvalPack datasets during the pre-training stage, as they provide balanced instances for various programming languages. We perform inference with all three models and collect latent representations for three widely-used programming languages: Python, C++, and Java.

For each (language, dataset) pair, we calculate the mean activation across instances and measure the distribution differences using the Wasserstein distance, where higher values indicate greater dissimilarity between distributions. We first compute the distance between programming languages within each dataset and then calculate the distance across datasets for the same programming language. The resulting distance matrix is illustrated in Figure~\ref{fig:rq3:distance}. We can observe a really intriguing phenomenon: the cross-task distance for the same language (diagonal entries) is generally larger than the in-task distance between different languages (low triangle and upper triangle). The only exception to this pattern is the Python v.s. Java pair for the {\codellama} model on the EvalPack dataset. Upon closer examination, we find that this anomaly may arise because, in the EvalPack dataset, only Java code begins with ``class solution'', whereas other programming languages use function-based answers. {\codellama} appears to be sensitive to such structural differences, leading to greater internal state variations for this specific case. Overall, these results suggest that the latent representations learned by the SAE can, to some extent, capture high-level task-related information.

\subsubsection{Feature Interpretability} 

\begin{figure}
    \includegraphics[width=0.45\textwidth]{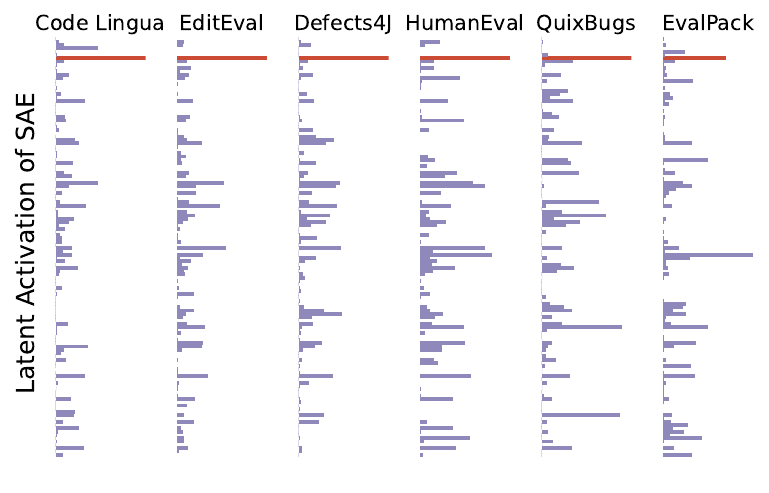}
    \vspace{-4mm}
    \caption{Activation difference between buggy and correct code lines on {\codellama}. Latent 121 is highlighted.}
    \label{fig:rq3_codellama}
\end{figure}

\begin{figure}
    \includegraphics[width=0.45\textwidth]{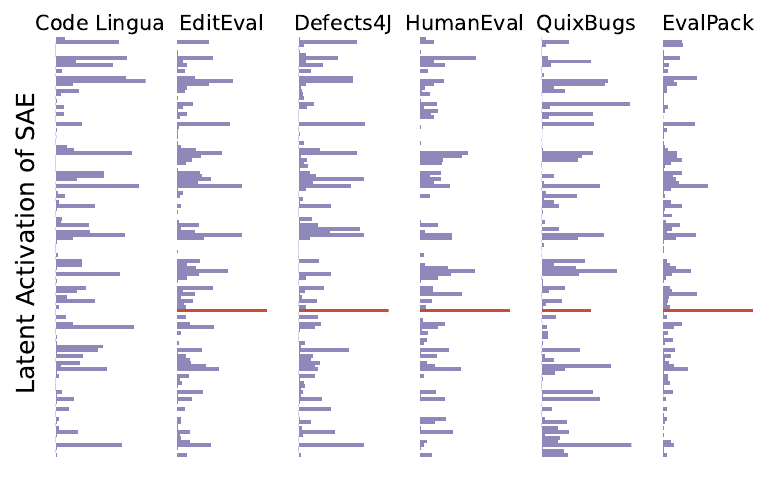}
    \vspace{-4mm}
    \caption{Activation difference between buggy and correct code lines on {\starcoder}. Latent 44 is highlighted.}
    \label{fig:rq3_starcoder}
    \vspace{-2mm}
\end{figure}

\begin{figure}
    \includegraphics[width=0.45\textwidth]{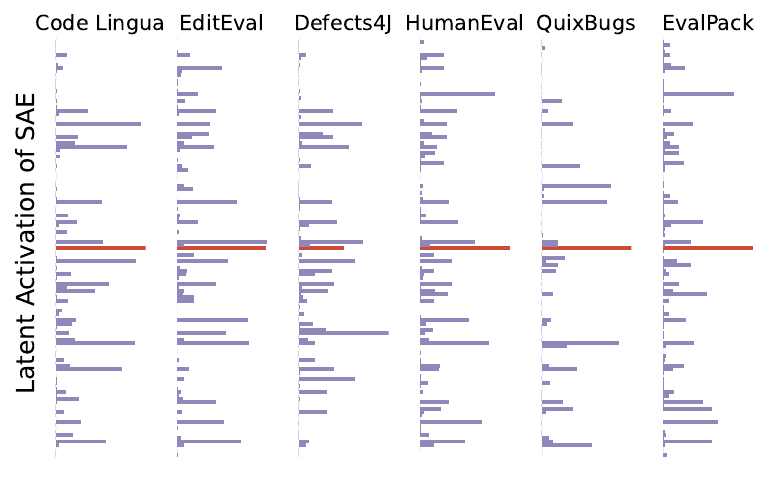}
    \vspace{-4mm}
    \caption{Activation difference between buggy and correct code lines on {\qwen}. Latent 63 is highlighted.}
    \label{fig:rq3_qwen}
    \vspace{-5mm}
\end{figure}

While {\our} allows replacing SAE with a more powerful autoencoder for potentially more accurate risk assessment, we select SAE because it offers a unique advantage: improved interpretability. Enabling users to observe and interact with a complex AI system fosters greater trust compared to providing a single risk score. This enhanced transparency can lead to significant advantages in real-world interactions between code LLMs and developers.

The interpretability of SAE comes from its ability to extract monosemantic features. The neurons of general DNNs have a phenomenon known as Superposition~\cite{elhage2022toy}, where models tend to represent more features than the neurons they have, and this leads to polysemanticity -- each neuron can convey different information under different contexts. This is good for performance but bad for interpretability. In contrast, SAE is found to be able to identify monosemantic features~\cite{bricken2023monosemanticity} and can better align with human understanding. Next, we present a simple showcase of this feature.

Given all the datasets in the semantic binding stage and evaluation stage (i.e., HumanEval, QuixBugs, Evalpack, {\translation}, {\edit}, and {\repair}), we select those that LLMs' answers that are incorrect. Within these samples, we calculate the average latent activation of the SAE for both incorrect and correct code lines, then subtract the latter from the former. Positive values in the resulting activation map highlight latents that are more activated in incorrect code lines compared to correct ones. We further average this activation map at the dataset level and visualize the results in Figure~\ref{fig:rq3_codellama}-\ref{fig:rq3_qwen}.

Surprisingly, we observe obvious cross-task latents across all studied code LLMs that represent erroneous code lines. It is important to note that these datasets encompass a wide range of tasks, including code editing, generation, translation, and repair, across both Python and Java programming languages. This finding may explain why {\our} demonstrates robust performance in risk assessment across diverse tasks. This behavior closely resembles Error-Related Negativity (ERN)~\cite{frank2005error} observed in human brains, where neurons exhibit clear signals when recognizing errors.

\begin{finding}
\label{finding:3}
    The SAE component in {\our} demonstrates robust representations that are consistent across both languages and tasks. Additionally, we identify intuitive features strongly associated with code LLMs' errors. This level of interpretability positions {\our} as a promising step toward establishing a systematic and trustworthy framework for code LLMs.
\end{finding}

\section{Discussion}

\subsection{Lesson Learned}

\noindent \textbf{Disentangle Representation and Computation.} Risk assessment for LLMs can be pretty challenging since they are trained on large-scale data. These data are usually inaccessible. Even when accessible, directly linking these data to the risk of LLMs is difficult because LLMs are optimized for next-token prediction, a paradigm significantly different from that of classical DNNs explicitly trained for specific functionalities.

The way we address this challenge is to disentangle representation from computation. Specifically, we train a feature extractor, the Sparse Autoencoder (SAE), in an unsupervised pre-training manner to capture the internal representations of the models’ states. Following this, we establish a mapping between the internal states and the correctness of the outputs through a supervised semantic binding process. This two-stage approach proves highly effective, especially when pre-training data are abundant but labeled data for downstream tasks are scarce.


\noindent \responseline{\textbf{The Critical Role of Data Management.} In the initial phase of our project, we concentrated primarily on developing pipeline prototypes, dedicating significant time to state extraction and SAE training. However, once we successfully executed a basic ``hello world'' example, the complexity of data management became apparent, eventually consuming the majority of our efforts. The challenge lay in handling various dataset versions across multiple tasks and programming languages, each with distinct pre- and post-processing requirements.}

\responseline{Despite these challenges, we discovered that enhancing data quality through methods such as adding datasets and refining the annotation/processing strategies proved more beneficial than merely optimizing training algorithms or adopting new architectures. This is particularly relevant in the era of LLMs, where data pipelines are more dynamic compared to the pre-LLM era, largely due to LLMs' multitasking capabilities. Consequently, we advocate for increased focus on the design and management of data pipelines among related stakeholders.}


\vspace{-2mm}

\subsection{Future Direction}

\noindent \textbf{Vast space yet to be explored.} In this study, we focus on probing a specific early layer of LLMs, as prior research suggests that these layers exhibit a stronger correlation with high-level decisions made by the models~\cite{zou2023representation, he2024llm, huang2024active}. However, many other layers remain untested and unexplored. Different layers might respond uniquely to different features—for instance, some may be more aligned with code performance, while others could be tied to code security. Furthermore, {\our} is not limited to analyzing hidden layers; it can also be extended to directly analyze attention layers, offering additional insights into the model’s internal mechanisms. Future work could take several directions. One approach is to determine, for a specific task (e.g., code line error detection), which layer contributes most effectively to final accuracy and how to identify it automatically. Another is to deepen our understanding of the roles played by individual layers, providing valuable guidance for subsequent research and practical applications.

\vspace{2mm}

\noindent \textbf{More trustworthiness properties.} In this work, we focus on applying {\our} to code line errors and code snippet errors. However, the semantic binding stage of {\our} is not explicitly restricted to specific types of labels, which means it could, in theory, be extended to address other properties, such as the security and performance of generated code, as long as the internal states contain relevant information. Extending {\our} to these applications would be an intriguing avenue for future research. Along the way, two key components could be further enhanced to support such different scenarios: (1) Refine the methods for extracting hidden states that contain richer and more task-relevant information, such as by enhancing the plain and contrastive loss functions at the pre-training stage. (2) Develop more effective ways to align the desired semantic properties with the learned representations, ensuring that the model captures the target characteristics more accurately at the semantic binding stage.

\vspace{2mm}

\noindent \textbf{Evaluation of the learned representation quality.} {\our} relies on the learned latent representations from the pre-training stage to achieve cross-language and cross-task effectiveness. However, before evaluating the entire framework on downstream tasks, it is challenging to directly assess the quality of the internal states. Currently, this quality can only be evaluated indirectly through downstream performance or manual feature analysis, as demonstrated in Section~\ref{section:result:rq3}. A promising direction for future work is to develop methods for assessing the quality of latent representations without relying on external labels. Such evaluation techniques could significantly accelerate the iterative improvement of the first-stage pre-training method, facilitate SAE comparisons, and further enhance the overall effectiveness of {\our}.

\vspace{2mm}

\section{Threats to Validity}

\noindent\textbf{Internal and Construct Validity.} One primary concern is the process of obtaining fine-grained line-level error labels, which are critical for both the semantic binding and evaluation stages. This process could potentially threaten internal and construct validity, as the GPT-4o models used at this stage may introduce errors and uncertainty.

To address this issue, we take a two-step approach. First, we instruct GPT-4o to correct the identified error lines in the given code and run automatic testing based on the provided test cases of each dataset. For cases where GPT-4o fails to give a correct answer, we conduct sampling and manual review to verify and correct the labels. In future work, we could enhance this process by directly utilizing fully human-labeled data collected from feedback platforms, further improving the reliability of the labels.

\vspace{2mm}

\noindent\textbf{External Validity.} The external validity of this study relates to the generalizability of our findings across different LLMs and tasks. To address this concern, we evaluate {\our} using three advanced LLMs from both industry and academia. Additionally, we include a diverse range of tasks—such as code editing, code translation, and code repair—across two programming languages in evaluation. Future work could consider incorporating a wider variety of code LLMs and expanding the range of tasks covered in the evaluation.

\section{Conclusion}

In this paper, we propose a pre-training framework {\our} for code LLMs that is, by its design, able to scale to vast amounts of training data. We outline a workflow for constructing {\our}, which can be seamlessly integrated into the existing infrastructure of software companies. Through systematic evaluation, we demonstrate the promising performance of {\our}, even with a very limited number of data available at the semantic binding stage. Our experiments show that the representations learned during the pre-training stage enable {\our} to generalize effectively to diverse software engineering tasks across different programming languages that the downstream classifier has never encountered during training. Additionally, {\our} provides intuitive and interpretable features, allowing users to better understand the potential risks in the answers generated by LLMs. Building on these findings, we call for related stakeholders to push the pre-training-based LLMs' risk assessment forward, paving the way for more robust and trustworthy LLM-driven systems.

\begin{acks}
\responseline{This work was supported by Fujitsu Limited.} 
\end{acks}

\balance

\bibliographystyle{ACM-Reference-Format}
\bibliography{citation.bib}

\end{document}